\documentclass[prb,letterpaper,aps,floatfix,twocolumn]{revtex4-1}
\usepackage{graphicx}
\usepackage{amsmath}
\usepackage{subfigure}
\newcommand{\beq}{\begin{equation}}
\newcommand{\eeq}{\end{equation}}
\newcommand{\bk}{{{\bf{k}}}}

\newcommand{\bK}{{{\bf{K}}}}

\newcommand{\br}{{{\bf{r}}}}

\newcommand{\bG}{{{\bf{G}}}}
\newcommand{\bA}{{\bf{A}}}

\newcommand{\bq}{{\bf{q}}}

\newcommand{\bb}{{\bf{b}}}

\newcommand{\bj}{{\bf j}}
\newcommand{\beqa}{\begin{eqnarray}}
\newcommand{\eeqa}{\end{eqnarray}}
\newcommand{\pdg}{{\vphantom \dag}}
\newcommand{\dg}{{\dag}}
\newcommand{\bnabla}{{\boldsymbol \nabla}} 
\newcommand{\bsigma}{{\boldsymbol \sigma}}
\newcommand{\btau}{{\boldsymbol \tau}}

\newcommand{\upa}{\uparrow}
\newcommand{\da}{\downarrow}

\newcommand{\cD}{{\cal D}}
\begin{document}
\title{Negative longitudinal magnetoresistance in Dirac and Weyl metals}
\author{A.A. Burkov}
\affiliation{Department of Physics and Astronomy, University of Waterloo, Waterloo, Ontario 
N2L 3G1, Canada} 
\affiliation{National Research University ITMO, Saint Petersburg 197101, Russia}
\date{\today}
\begin{abstract}
It has recently been found that Dirac and Weyl metals are characterized by an unusual weak-field longitudinal magnetoresistance: large, negative, and quadratic in the 
magnetic field. This has been shown to arise from chiral anomaly, i.e. nonconservation of the chiral charge in the presence of external electric and magnetic fields, 
oriented collinearly. In this paper we report on a theory of this effect in both Dirac and Weyl metals. We demonstrate that this phenomenon contains two important 
ingredients. One is the magnetic-field-induced coupling between the chiral and the total (or vector, in relativistic field theory terminology) charge densities. This arises 
from the Berry curvature and is present in principle whenever the Berry curvature is nonzero, i.e. is nonspecific to Dirac and Weyl metals. This coupling, however, leads to a large 
negative quadratic magnetoresistance only when the second ingredient is present, namely when the chiral charge density is a nearly conserved quantity with a 
long relaxation time. This property is specific to Dirac and Weyl metals and is realized only when the Fermi energy is close to Dirac or Weyl nodes, expressing an important 
low-energy property of these materials, emergent chiral symmetry. 
\end{abstract}
\maketitle
\section{Introduction}
\label{sec:1}
Weyl and closely related Dirac semimetals are the most recent addition to the growing family of materials with topologically-nontrivial electronic structure.
Both were initially predicted theoretically,~\cite{Wan11,Ran11,Burkov11-1,Xu11,Kane12,Fang12,Fang13,Bernevig15-1,Hasan15-5} and realized 
very recently in a remarkable series of experiments.~\cite{Hasan15-3,Fang15-2,Hasan15-4,Hasan15-2,Lu15,Fang15-1,Hasan15-1,Borisenko14,Shen14,Neupane14,Hasan15-6,Ando11,Kharzeev14,Gu15}
What distinguishes Weyl semimetals from other topologically nontrivial states of matter, like topological insulators (TI),~\cite{Hasan10,Qi11} is that they are gapless.
The topological object in this case is a point of contact between two nondegenerate bands at the Fermi level, which acts as a monopole source of Berry curvature 
and thus carries an integer topological charge. 
The significance of such electronic structure features was emphasized in earlier pioneering work of Volovik,~\cite{Volovik03,Volovik07} which partly anticipated the recent developments.
 
A hallmark of ``topological" states of matter is the presence of metallic edge states, which arise necessarily due to the impossibility of a smooth connection between
the topologically nontrivial sample and its topologically trivial environment. 
Weyl semimetal does indeed possess such surface states, whose topologically nontrivial nature is manifest in the shape of their Fermi surface, having the form of an open 
arc (Fermi arc), rather than a closed curve, as in any regular two-dimensional (2D) metal.  These have been seen directly using ARPES in the newly discovered 
Weyl semimetal materials TaAs and NbAs.~\cite{Hasan15-3,Hasan15-4,Fang15-2,Hasan15-6}

However, topologically nontrivial phases of matter often also have unusual electromagnetic response, the most famous example being the precisely quantized 
transverse conductivity of a 2D quantum Hall liquid. 
Such a response is a robust, detail-independent manifestation of the nontrivial electronic structure topology on macroscopic scales and is thus of particular interest. 

Topological electromagnetic response may be conveniently expressed as a topological term, generated in the action of the electromagnetic field, when electrons 
in the occupied states are integrated out.  
In the case of the 2D quantum Hall liquid, this is the Chern-Simons term
\beq
\label{eq:1}
S = -\frac{e^2}{4 \pi} \int d t \, d^2 r \, \epsilon^{\nu \alpha \beta} A_{\nu} \partial_{\alpha} A_{\beta}, 
\eeq
where (and henceforth) $\hbar = c = 1$ units are used, and filling factor one is taken above for concreteness.  
This may be generalized to 3D by stacking the 2D quantum Hall systems along a particular spatial direction,~\cite{Kohmoto92} which gives
\beq
\label{eq:2}
S = -\frac{e^2}{8 \pi^2} \int d t \,  d^3 r \, G_{\mu} \epsilon^{\mu \nu \alpha \beta} A_{\nu} \partial_{\alpha} A_{\beta},
\eeq
where $\bG = 2 \pi \hat n / d$ is a reciprocal lattice vector in the stacking direction $\hat n$, corresponding to superlattice period $d$. 
Using integration by parts, Eq.~\eqref{eq:2} may be rewritten in the following form, which will prove useful below 
\beq
\label{eq:3}
S = \frac{e^2}{32 \pi^2} \int d t \, d^3 r \, \theta(\br)\,  \epsilon^{\mu \nu \alpha \beta} F_{\mu \nu} F_{\alpha \beta}, 
\eeq
where $\theta(\br) = \bG \cdot \br$. 

For a 3D TI, the topological term takes the form of the so-called $\theta$-term~\cite{Qi11}
\beq
\label{eq:4}
S = \frac{e^2}{32 \pi^2} \int d t\, d^3 r \, \theta \, \epsilon^{\mu \nu \alpha \beta} F_{\mu \nu} F_{\alpha \beta}, 
\eeq
where $\theta = \pi$ for a TI and $\theta = 0$ for a normal insulator (NI). 
Applying integration by parts as above, it is clear that Eq.~\eqref{eq:4} in fact has no observable consequences in the bulk of the TI, since it is a 
total derivative and vanishes upon integration by parts. It does have an effect on the TI boundaries: upon breaking time-reversal (TR) symmetry
by e.g. magnetic impurity doping, it leads to half-quantized anomalous Hall effect (AHE).~\cite{Hasan10,Qi11} 

Weyl semimetal may be regarded as an intermediate phase between a 3D TI (or NI) and the 3D quantum Hall insulator, described by Eq.~\eqref{eq:3}.~\cite{Burkov11-1}
The topological term of a Weyl semimetal takes the form of Eq.~\eqref{eq:3}, but with the field $\theta$ given by~\cite{Zyuzin12-1,Hughes14}
\beq
\label{eq:5}
\theta(\br, t) = 2 \bb \cdot \br - 2 b_0 t, 
\eeq
where 
\beq
\label{eq:6}
\bb = \frac{1}{2} \sum_i C_i \bK_i, \,\, b_0 = \frac{1}{2} \sum_i C_i \epsilon_i. 
\eeq
Here $C_i$ is the topological charge of the $i$th Weyl node, $\bK_i$ is its location in momentum space and $\epsilon_i$ is its energy. 
This is similar to topological terms arising in the context of Lorentz-invariance-violating extensions of the Sandard Model of particle physics.~\cite{Klinkhamer05}
As is easy to see, the linear space-time coordinate dependence of the $\theta(\br, t)$ field is the only nontrivial dependence, compatible with space-time translational 
symmetry. 
Due to this space-time coordinate dependence, the topological term in Weyl semimetals does not vanish upon integration by parts, but instead takes the form, similar to 
Eq.~\eqref{eq:2}
\beq
\label{eq:7}
S = - \frac{e^2}{8 \pi^2} \int d t \, d^3 r \, \partial_{\mu} \, \theta(\br, t) \epsilon^{\mu \nu \alpha \beta} A_{\nu} \partial_{\alpha} A_{\beta}. 
\eeq
This, in turn, leads to a nontrivial modification of the Maxwell equations in the bulk of the Weyl semimetal, which may be expressed as two extra contributions 
to the current density, obtained by varying Eq.~\eqref{eq:7} with respect to the electromagnetic gauge potential
\beq
\label{eq:8}
j_{\nu} = \frac{e^2}{2 \pi^2} b_{\mu} \epsilon^{\mu \nu \alpha \beta} \partial_{\alpha} A_{\beta}, \,\, \mu = 1, 2, 3, 
\eeq
and 
\beq
\label{eq:9}
j_{\nu} = - \frac{e^2}{2 \pi^2} b_0 \epsilon^{0 \nu \alpha \beta} \partial_{\alpha} A_{\beta}. 
\eeq
Eq.~\eqref{eq:8} describes AHE with semi-quantized Hall conductivity, proportional to the magnitude of the vector $\bb$, giving the separation between 
the Weyl nodes in momentum space, while Eq.~\eqref{eq:9} describes the so-called chiral magnetic effect (CME).~\cite{Kharzeev08}
We will return to the meaning of the latter equation below. 

Topological term, describing the electromagnetic response of Weyl semimetals (and 3D TI as well), may be regarded as being a consequence of 
chiral anomaly,~\cite{Adler69,Jackiw69,Nielsen83,Aji12,Zyuzin12-1,Hosur10} a fundamentally important concept in relativistic field theory,
which has recently found its way into condensed matter physics and plays an important role in the modern understanding of topologically-nontrivial 
phases of matter.~\cite{Ludwig12,Furusaki13}
However, when using field theory concepts in the condensed matter context, one needs to exercise some care. Relativistic field theories possess
exact symmetries, which in condensed matter systems may only be approximate.  
In particular, chiral anomaly is closely related to chiral symmetry, i.e. separate conservation of fermions of left and right chirality, which is an exact symmetry in theories 
of massless relativistic particles. Chiral anomaly refers to violation of this symmetry by quantum effects in the presence of electromagnetic field. 
In real Weyl or Dirac semimetals this symmetry may exist only approximately, if the Fermi energy is sufficiently close to the location of the nodes, so that the band dispersion may be 
taken to be linear to a good approximation. It is then unclear to what extent the concept of chiral anomaly is meaningful, when applied to Weyl and Dirac semimetals. 
 
In fact, the first sign of trouble with Eq.~\eqref{eq:7} is its immediate consequence, Eq.~\eqref{eq:9}, which describes the CME. This equation has the appearance of a current, driven by an applied magnetic field in the presence of an energy separation between the nodes. 
Such an energy separation may exist in equilibrium in a noncentrosymmetric material,~\cite{Zyuzin12-2} which leads one to a problematic conclusion that Eq.~\eqref{eq:9} describes an 
equilibrium current. This would violate basic principles of condensed matter physics and can not happen.~\cite{Vazifeh13}
The origin of this problem is precisely the relativistic invariance, assumed in the derivation of Eq.~\eqref{eq:7},~\cite{Zyuzin12-1} but not actually present in a real Weyl or Dirac 
semimetal. 
As was shown in Ref.~\onlinecite{Chen13-2}, in a condensed matter setting, the response, described by Eq.~\eqref{eq:9}, depends on the order, in which the limits of zero frequency 
and zero wavevector are taken. When the frequency is taken to zero before the wavevector is taken to zero, which corresponds to thermodynamic equilibrium response, the current vanishes. However, when the order of limits is reversed, which corresponds to the DC limit of  nonequilibrium response, the current is nonzero and given by Eq.~\eqref{eq:9}. 
This dependence on the order of limits appears to violate Lorentz invariance and thus should not happen in a relativistic particle physics context (at least if Lorentz invariance is assumed 
to be a fundamental symmetry). This highlights the importance of being careful when using low-energy models, exhibiting ``relativistic" properties, to describe Weyl semimetals 
and other Dirac materials.  

In this paper we will describe observable effects of the chiral anomaly, in particular the observable manifestation of CME, in model Dirac and Weyl metals, i.e. lightly doped 
Dirac and Weyl semimetals. In accordance with the discussion above, we will use models for both, which explicitly do not possess chiral symmetry and are thus free of the artifacts 
of ``relativistic" low energy models. 
We will demonstrate that in both Dirac and Weyl metals the main experimentally-observable consequence of CME is an unusual weak-field longitudinal magnetoresistance, 
which is negative, quadratic in the magnetic field, and large when the Fermi energy is sufficiently close to the Dirac or Weyl nodes.~\cite{Spivak12} A shorter account of this work, devoted specifically to the Weyl metal case, has already been published.~\cite{Burkov14-3,Burkov15-1}
The rest of the paper is organized as follows. 
In Section~\ref{sec:2} we introduce the model we will use to describe both Dirac and Weyl metals, and which is based on the TI-NI heterostructure model of Weyl semimetals, 
introduced by us before.~\cite{Burkov11-1} This model is the simplest model of a Dirac or Weyl metal that does not suffer from the ``relativistic" artifacts, in particular it does 
not possess the spurious chiral symmetry of low-energy models with purely linear dispersion. 
In Section~\ref{sec:3} we describe how CME manifests in Dirac metals. We derive coupled transport equations for, using field theory terminology, the vector and the axial (chiral) charge
densities, which are coupled in the presence of an applied magnetic field. The coupling is shown to be the manifestation of CME. However, we demonstrate that this only leads to 
experimentally measurable consequences when a second ingredient is present: near conservation of the axial charge density, which is never exact, but becomes more and 
more precise as the Fermi energy approaches the Dirac node. 
In Section~\ref{sec:4} we describe the same effect in Weyl semimetals. The manifestation of CME in Weyl semimetals is found to be nearly identical to the Dirac semimetals, i.e. whether the nodes are separated in momentum space or not does not matter for this effect. This appears to not be fully appreciated in the literature. 
We conclude in Section~\ref{sec:5} with a brief discussion of our main results and experimental observability of the effect. 
\section{Model and preliminaries}
\label{sec:2}
We start from a model of Weyl and Dirac metals, based on TI-NI multilayer heterostructure, introduced by us.~\cite{Burkov11-1} 
The advantage of this model is that it is extremely simple, yet more realistic than the most generic low-energy model of a Dirac or Weyl metals would be, 
in particular it does not have the unphysical chiral symmetry. 
Since the model has already been described in a number of publications, here we will only recap the most essential points. 
The momentum space Hamiltonian, describing the multilayer structure, is given by
\beq
\label{eq:10}
H = v _F \tau^z (\hat z \times \bsigma) \cdot \bk + \hat \Delta(k_z). 
\eeq
Here $\hat z$ is the growth direction of the heterostructure, $v_F$ is the Fermi velocity, associated with the motion in the transverse ($x, y$) directions,
$\bsigma$ are Pauli matrices, describing the real spin degree of freedom, while $\btau$ is the pseudospin, describing the top and bottom surfaces of TI layers
in the heterostructure. The operator $\hat \Delta(k_z)$ describes the electron dynamics in the growth direction and is explicitly given by
\beq
\label{eq:11}
\hat \Delta(k_z) = \Delta_S \tau^x + \frac{\Delta_D}{2} \left( \tau^+ e^{i k_z d} + h.c. \right), 
\eeq
where $\Delta_{S,D}$ are amplitudes for tunnelling between top and bottom surfaces of the same (S) or neighbouring (D) TI layers and $d$ is the superlattice period. 
We will take them both to be positive for concreteness. 
This structure is an ordinary insulator when $\Delta_S > \Delta_D$ and a strong 3D TI otherwise. 
The point $\Delta_S = \Delta_D$ marks the TI-NI phase transition. At this point the structure is a Dirac semimetal.
Weyl semimetal is obtained by adding a TR breaking term $b \sigma^z$, which arises physically either from polarized magnetic impurities or an 
external magnetic field. This has been described in our earlier papers,~\cite{Burkov11-1,Burkov11-2} and we will not dwell on it further. 

To make contact with chiral anomaly, it is useful to recast Eq.~\eqref{eq:10} in a ``relativistic" form. To this end we expand Eq.~\eqref{eq:10} to leading 
order in the crystal momentum near the 
point $\bk = (0, 0, \pi/d)$, at which the gap closing occurs at the TI-NI transition when $\Delta_S = \Delta_D$.
We obtain
\beq
\label{eq:12}
H = v _F \tau^z (\hat z \times \bsigma) \cdot \bk + \Delta_D d \tau^y k_z + \left(\Delta_S - \Delta_D \right) \tau^x.
\eeq
This implies the following representation of the first four Dirac gamma matrices
\beq
\label{eq:13}
\gamma^0 = \tau^x, \,\, \gamma^1 = i \tau^y \sigma^y, \,\, \gamma^2 = - i \tau^y \sigma^x, \,\, \gamma^3 = i \tau^z. 
\eeq
The fifth gamma matrix 
\beq
\label{eq:14}
\gamma^5 = i \gamma^0 \gamma^1 \gamma^2 \gamma^3 = \tau^y \sigma^z, 
\eeq
defines the axial charge operator $n_a = \gamma^5 = \tau^y \sigma^z$. 
Absorbing the Fermi velocities $v_F$ and $\tilde v_F = \Delta_D d$ into the definition of the corresponding momentum components and 
replacing $k^{\mu} \rightarrow - i \partial_{\mu}$, one obtains the following ``relativistic" Lagrangian
\beq
\label{eq:15}
{\cal L} = \psi^\dg i \partial_t \psi^\pdg - H = \bar \psi\,\left[ i \gamma^{\mu}(\partial_{\mu} + i e A_{\mu} + i b_{\mu} \gamma^5) - m \right] \psi.
\eeq
Here $m = \Delta_S - \Delta_D$ is the Dirac ``mass", $\bar \psi = \psi^\dg \gamma^0$ is the Dirac adjoint of the Grassmann field $\psi$, $A_{\mu}$ is the electromagnetic gauge potential, and we have 
also introduced chiral gauge field $b_{\mu}$. Explicitly, chiral gauge field arises from the following terms, added to the Hamiltonian Eq.~\eqref{eq:12}
\beq
\label{eq:16}
H_b = b_0 \tau^y \sigma^z + b_1 \tau^x \sigma^x + b_2 \tau^x \sigma^y + b_3 \sigma^z. 
\eeq
The first term in Eq.~\eqref{eq:16} is clearly an axial chemical potential term, which shifts the left (L) and right-handed (R) components of the Dirac fermion 
in opposite directions in energy. The last term is magnetization (or magnetic field) in the $z$-direction, which shifts the L and R components in opposite directions 
in momentum space along the $z$-axis. 
The second and third terms have the same symmetry as magnetization components in the $x,y$ directions, and thus may be regarded as such. However, one 
needs to be aware that bare $\sigma^{x,y}$ operators will have a very different effect on the spectrum, creating a nodal line state rather than point nodes.~\cite{Burkov11-2}

Chiral anomaly refers to anomalous nonconservation of the axial current $J^{\mu}_5 = \bar \psi \gamma^{\mu} \gamma^5 \psi$. 
This means that the axial current continuity equation has the following form
\beq
\label{eq:17}
\partial_{\mu} J^{\mu}_5 = 2 i m \, \bar \psi \gamma^5 \psi + \frac{e^2}{16 \pi^2} \epsilon^{\mu \nu \alpha \beta} F_{\mu \nu} F_{\alpha \beta}.
\eeq
The first term on the right hand side of Eq.~\eqref{eq:17} is the classical contribution to the axial charge continuity equation, which is easily obtained from the 
Dirac equation. When the mass $m = 0$, i.e. when $\Delta_S = \Delta_D$, this term vanishes. This is an expression of chiral symmetry, i.e. classically the axial 
charge is conserved when $m = 0$. This conservation is violated when the continuity equation is evaluated in the second-quantized theory, which is where 
the second term comes from. 
However, $m = 0$ when $\Delta_S = \Delta_D$ is only obtained at leading order in the expansion of the operator $\hat \Delta(k_z)$ near $k_z = \pi/d$. 
In fact, $m$ is a function of $k_z$ and does not generally vanish. This means that the chiral symmetry is always only approximate and the axial charge 
in never a truly conserved quantity, even when the anomaly term is neglected. 
It is then clear that anomaly-related effects may only be observable if chiral symmetry is almost there, i.e. the axial charge relaxation time is long. 
We generally expect it to be long when the Fermi energy is close to zero, and thus the effect of higher order terms in the expansion of $\hat \Delta(k_z)$ is
not significant. The purpose of the rest of the paper is to make these statements quantitative and evaluate the axial relaxation time explicitly for model Dirac 
and Weyl metals, described by Eq.~\eqref{eq:10}. 

\section{Anomalous density response in a Dirac metal}
\label{sec:3}
In this section, starting from a microscopic model of a Dirac semimetal, given by Eq.~\eqref{eq:10}, we will derive transport equations for the vector and axial charge 
densities, in the presence of an external magnetic field. 
As will be demonstrated, chiral anomaly will manifest in these equations as a coupling between the vector and the axial charge densities, induced by the magnetic field. 
We will discuss under what conditions this coupling leads to observable transport phenomena, namely quadratic negative magnetoresistance. 

We start from the Hamiltonian Eq.~\eqref{eq:10}, with an added magnetic field in the $z$-direction
\beq
\label{eq:18}
H = v_F \tau^z (\hat z \times \bsigma) \cdot (- i \bnabla + e \bA) + \hat \Delta(k_z).
\eeq
We will adopt Landau gauge for the vector potential $\bA = x B \hat y$ and ignore the Zeeman splitting in this section. The restriction of the magnetic field to the $z$-direction 
simplifies calculations
in the context of our model, but is otherwise nonessential. 
After a canonical transformation 
\beq
\label{eq:18.5}
\sigma^{\pm} \rightarrow \tau^z \sigma^{\pm}, \,\, \tau^{\pm} \rightarrow \sigma^z \tau^{\pm}, 
\eeq
Eq.~\eqref{eq:18} is easily diagonalized. 
The eigenvalues have the form
\beq
\label{eq:19}
\epsilon_{n a}(k_z) = s \sqrt{2 \omega_B^2 n + \Delta^2(k_z)} \equiv s \epsilon_n(k_z).
\eeq
Here $n \geq1$ is the main Landau level (LL) index, $k_y$ is the intra-LL orbital label,  $s = \pm$ labels the two sets of positive and negative energy (or electron-like and hole-like) eigenvalues, $\omega_B = v_F /\ell_B$ is the Dirac cyclotron frequency and 
$\ell_B = 1/\sqrt{e B}$ is the magnetic length. The index $a$ is a composite index $a = s, t$, where $t = \pm$ labels two components of the Kramers doublet. The energy eigenvalues 
do not depend on $t$ as we have ignored the Zeeman splitting due to the applied field. 
$t \Delta(k_z)$ are the two eigenvalues of the $\hat \Delta(k_z)$ operator, where $\Delta(k_z) = \sqrt{\Delta_S^2 + \Delta_D^2 + 2 \Delta_S \Delta_D \cos(k_z d)}$. 
The corresponding eigenstates have the following form
\beqa
\label{eq:20}
|n, a, k_y, k_z \rangle&=&\sum_{\tau} \left[ z^a_{n \upa \tau} (k_z) |n-1, k_y, k_z, \upa, \tau \rangle \right. \nonumber \\
&+&\left. z^a_{n \da \tau}(k_z) |n, k_y, k_z, \da, \tau \rangle \right], 
\eeqa
where
\beq
\label{eq:21}
\langle \br | n, k_y, k_z, \sigma, \tau \rangle = \frac{1}{\sqrt{L_z}} e^{i k_z z} \phi_{n k_y}(\br) | \sigma, \tau \rangle, 
\eeq
$\phi_{n k_y}(\br)$ are the Landau-gauge orbital wavefunctions and $\sigma, \tau$ are the spin and the top-bottom surface pseudospin labels
respectively. 
The four-component eigenvector $| z^a_{n}(k_z) \rangle$ may be written as a tensor product of the two-component spin and pseudospin eigenvectors, 
i.e. $| z^a_{n}(k_z) \rangle = | v^a_{n}(k_z) \rangle \otimes | u^a(k_z) \rangle$, where 
\beqa
\label{eq:22}
&&|v^{s t}_{n}(k_z) \rangle = \frac{1}{\sqrt{2}} \left(\sqrt{1 + s \frac{t \Delta(k_z)}{\epsilon_{n}(k_z)}}, - i s \sqrt{1 - s \frac{t \Delta(k_z)}{\epsilon_{n}(k_z)}} \right), \nonumber \\
&&|u^t(k_z) \rangle = \frac{1}{\sqrt{2}} \left(1, t \frac{\Delta_S + \Delta_D e^{- i k_z d}}{\Delta(k_z)} \right). 
\eeqa
The lowest $n = 0$ LL is special, which is a consequence of nontrivial Berry curvature. The $s$ quantum number is absent in this case and  
taking $B > 0$ for concreteness, we have 
\beq
\label{eq:23}
\epsilon_{n t}(k_z) = - t \Delta(k_z), 
\eeq
and $|v^t_{0}(k_z) \rangle = (0,1)$. 

We model the potential due to random impurities as Gaussian white noise potential with $\langle V(\br) \rangle = 0$ and 
\beq
\label{eq:24}
\langle V(\br) V (\br') \rangle = \gamma^2 \delta(\br - \br'). 
\eeq
For simplicity we assume that the impurity potential is independent of the pseudospin index $\tau$, which physically means that we ignore scattering between 
the top and botton surfaces of the TI layers. This does not affect the results qualitatively and is only done to simplify calculations. 

\begin{figure}[t]
  \includegraphics[width=11cm]{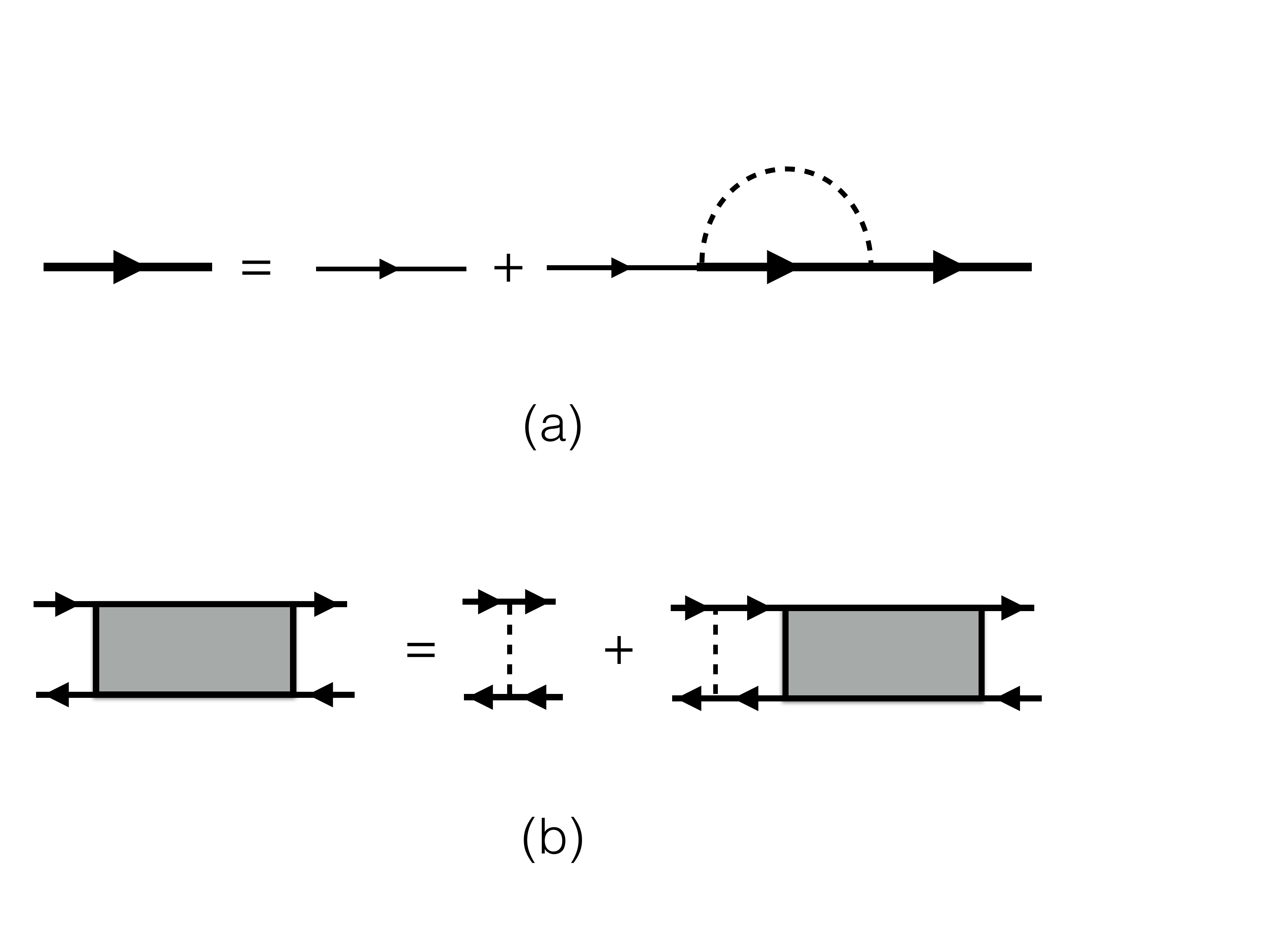}
  \caption{(a) Graphical representation of the SCBA equation. Regular lines denote bare Green's functions, while bold lines correspond to impurity-averaged Green's 
  functions. Dashed line denotes impurity averaging. (b) Graphical representation of the diffusion propagator, which is the sum of all ladder diagrams. Bold lines correspond to impurity-averaged Green's functions.}
  \label{fig:1}
\end{figure}

The impurity scattering may be treated by the standard diagrammatic perturbation theory. 
In the self-consistent Born approximation (SCBA) one obtains the following expression for the retarded 
impurity self energy (see Fig.~\ref{fig:1}~(a) for graphical representation)
\beqa
\label{eq:25}
\Sigma^R_{n a}(k_z, \omega)&=&\frac{1}{L_z} \sum_{n' a' k_y'  k_z'} \langle |\langle n  a  k_y k_z| V |n'  a'  k_y'  k_z'\rangle|^2 \rangle \nonumber \\
&\times& G^R_{n' a'}(k_z', \omega),
\eeqa
where 
\beq
\label{eq:26}
G^R_{n a} (k_z, \omega) = \frac{1}{\omega - \xi_{n a}(k_z) + i \eta}, 
\eeq
is the retarded Green's function of a clean Dirac metal (full self-consistency is unnecessary here) and $\xi_{n a}(k_z) = \epsilon_{n a}(k_z) - \epsilon_F$. 
We will assume that the Fermi energy $\epsilon_F > 0$, i.e. that the Dirac semimetal is electron-doped. 
Matrix elements of the impurity potential between the LL eigenstates are easily evaluated and are given by
\beqa
\label{eq:27}
&&\langle n a k_y k_z |V| n' a' k_y' k_z' \rangle = \frac{1}{L_x L_y L_z} \sum_{\bq} \nonumber \\ 
&\times&V(\bq) \delta_{q_y, k_y - k_y'} \delta_{q_z, k_z - k_z'} e^{i q_x \ell_B^2 (k_y + k_y')/2} \nonumber \\
&\times&\sum_{\tau} \left[\bar z^a_{ n \upa \tau}(k_z) z^{a'}_{n' \upa \tau}(k_z') F_{n-1, n'-1}(\bq) \right. \nonumber \\
&+& \left.\bar z^a_{ n \da \tau}(k_z) z^{a'}_{n'  \da \tau} (k_z')F_{n, n'}(\bq)\right], 
\eeqa  
where the LL formfactors have the following well-known form
\beqa
\label{eq:28}
F_{n, n'} (\bq)&=&\sqrt{\frac{n'!}{n!}} \left(\frac{i q_x \ell_B - q_y \ell_B}{2}\right)^{n - n'}  \nonumber \\
&\times&e^{-q^2 \ell_B^2/ 4} L^{n - n'}_{n'} \left(\frac{q^2 \ell_B^2}{2}\right), 
\eeqa
and $L^{n - n'}_{n'}(x)$ are the generalized Laguerre polynomials. 
Using the following properties of the LL formfactor momentum integrals
\beqa
\label{eq:29}
&&\int \frac{d^2 q}{(2 \pi)^2} F_{n,n'}(\bq) F_{n', n}(-\bq) = \frac{1}{2 \pi \ell_B^2}, \nonumber \\
&&\int \frac{d^2 q}{(2 \pi)^2} F_{n,n'}(\bq) F_{n'-1, n-1}(-\bq) = 0, 
\eeqa 
we obtain
\beqa
\label{eq:30}
&&\Gamma_{n a, n' a'}(k_z, k_z') \equiv \langle |\langle n  a  k_y  k_z| V |n' a'  k_y'  k_z'\rangle|^2 \rangle \nonumber \\
&=&\gamma^2 \left[|v^a_{n \upa}(k_z)|^2 |v^{a'}_{n' \upa}(k_z')|^2 + 
|v^a_{n \da}(k_z)|^2 |v^{a'}_{n' \da}(k_z')|^2 \right] \nonumber \\
&\times&| \langle u^a(k_z) | u^{a'}(k_z') \rangle |^2. 
\eeqa
The SCBA equation then takes the form
\beq
\label{eq:31}
\Sigma^R_{n a}(k_z, \omega) = \frac{1}{2 \pi \ell_B^2 L_z} \sum_{n' a' k_z'} \Gamma_{n a, n' a'}(k_z, k_z') G^R_{n' a'}(k_z', \omega). 
\eeq
At this point we will assume that $\epsilon_F$ is sufficiently large, so that scattering between the electron- and hole-like LL may be neglected. 
Then the negative energy $s = -$ states do not contribute and we will ignore them henceforth. 
We will also drop the explicit $s = +$ index, since all states are the $s = +$ states, and replace $a$ index by $t$ from now on. 

The SCBA equation may now be solved analytically. 
Using 
\beq
\label{eq:32}
\textrm{Im} G^R_{n t}(k_z, \omega) = - \pi \delta[\epsilon_n(k_z) - \epsilon_F], 
\eeq
it is easy to see that the dependence of the matrix element $\Gamma_{n t, n' t'}(k_z, k_z')$ on the LL indices $n, n'$ in fact drops out, since this dependence only 
enters through the LL energies $\epsilon_n(k_z)$, which may simply be replaced by the Fermi energy. 
We then obtain the following expression for the impurity scattering rate
\beqa
\label{eq:33}
&&\frac{1}{\tau(k_z)} \equiv - 2 \textrm{Im} \Sigma^R_{n t}(k_z, \omega) = \frac{1}{\tau_0}\left[1 + \frac{\Delta_S + \Delta_D \cos(k_z  d)}{\epsilon_F} \right. \nonumber \\
&\times&\left. \frac{\Delta_S + \Delta_D \langle \cos(k_z d) \rangle}{\epsilon_F} \right]. 
\eeqa
Here 
\beq
\label{eq:34}
\frac{1}{\tau_0} = \frac{1}{2} \pi \gamma^2 g(\epsilon_F),
\eeq
and $g(\epsilon_F)$ is the total density of states at Fermi energy. Explicitly
\beqa
\label{eq:35}
g(\epsilon_F)&=&\frac{1}{2 \pi \ell_B^2} \int_{-\pi/d}^{\pi/d} \frac{d k_z}{2 \pi} \sum_{n t} \delta[\epsilon_n(k_z) - \epsilon_F)] \nonumber \\
&=& \frac{\epsilon_F}{\pi v_F^2} \int_{-\pi/d}^{\pi/d} \frac{d k_z}{2 \pi} \Theta[\epsilon_F - \Delta(k_z)], 
\eeqa
where in the second line above we have assumed that the magnetic field is weak and converted the sum over the LL index $n$ to an integral. 
Finally, $\langle \cos(k_z d) \rangle$ in Eq.~\eqref{eq:33} means the average of $\cos(k_z d)$ over the Fermi surface, which is defined as
\beq
\label{eq:36}
\langle \cos(k_z d) \rangle = \frac{2}{g(\epsilon_F)} \int_{-\pi/d}^{\pi/d} \frac{d k_z}{2 \pi} \sum_n \cos(k_z d) \delta[\epsilon_n(k_z) - \epsilon_F)]. 
\eeq
We will use this definition of Fermi surface averages throughout the paper. 
Evaluating the average in Eq.~\eqref{eq:36} in the weak-field limit, one obtains
\beq
\label{eq:37}
\langle \cos(k_z d) \rangle = -\frac{1}{k_0 d} \sqrt{1- \left(\frac{\Delta_S^2 + \Delta_D^2 - \epsilon_F^2}{2 \Delta_S \Delta_D}\right)^2},
\eeq
where
\beq
\label{eq:38}
k_0 = \frac{1}{d}\arccos\left(\frac{\Delta_S^2 + \Delta_D^2 - \epsilon_F^2}{2 \Delta_S \Delta_D}\right), 
\eeq
is the solution of the equation $\Delta(k_z) = \epsilon_F$ modulo $\pi/d$.

At this point we will specialize to the case of a Dirac metal, i.e. set $\Delta_S = \Delta_D$, which makes the Dirac mass at the point $k_x = k_y = 0, k_z = \pi/d$ vanish.
We will also assume that the Fermi energy is close to the Dirac point, which means $\epsilon_F/\Delta_S \ll 1$ (but still far enough that $\epsilon_F \tau_0 \gg 1$). 
Then we obtain
\beq
\label{eq:39}
\langle \cos(k_z d) \rangle \approx -1 + \frac{\epsilon_F^2}{6 \Delta_S^2} + \frac{\epsilon_F^4}{180 \Delta_S^4} + \ldots, 
\eeq
and it is clear that, to leading order in the small parameter $\epsilon_F/ \Delta_S$,
\beq
\label{eq:40}
\frac{1}{\tau(k_z)} \approx \frac{1}{\tau_0}. 
\eeq
 
We now want to find the propagator of the diffusion modes in our system, which correspond to nearly conserved physical quantities with long (i.e. much longer than $\tau_0$) relaxation times. 
In the limit $\epsilon_F \tau_0 \gg 1$, the diffusion propagator may be evaluated by summing ladder impurity scattering diagrams, as shown in Fig.~\ref{fig:1}~(b).~\cite{Altland10}
This approximation (self-consistent non-crossing approximation) is consistent with the SCBA for the impurity self-energy, in the sense that together they preserve exact conservation laws, in particular charge conservation. 
The diffusion propagator (or diffuson), evaluated in the self-consistent non-crossing approximation, takes the following form
\beq
\label{eq:41}
{\cal D} (\bq, \Omega) = \left[1 - I(\bq, \Omega)\right]^{-1}, 
\eeq
where $I(\bq, \Omega)$ is a $16 \times 16$ matrix with respect to the combined spin and pseudospin indices $\alpha = \sigma, \tau$, which has the following explicit form 
\beqa
\label{eq:42}
&&I_{\alpha_1 \alpha_2, \alpha_3 \alpha_4}(\bq, \Omega) = \frac{\gamma^2}{L_x L_y L_z} \int d^3 r d^3 r' e^{-i \bq \cdot (\br - \br')} \nonumber \\
&\times&G^R_{\alpha_1 \alpha_3}(\br, \br'| \Omega) G^A_{\alpha_4 \alpha_2}(\br', \br | 0). 
\eeqa
The Green's functions, appearing in Eq.~\eqref{eq:42} are the impurity-averaged SCBA retarded and advanced Green's functions, found above
\beq
\label{eq:43}
G^{R,A}_{\alpha \alpha'}(\br, \br' |\Omega) = \sum_{n t k_y k_z} \frac{\langle \br  \alpha | n t k_y k_z \rangle \langle n t k_y k_z | \br'  \alpha' \rangle}
{\Omega - \xi_{n t} (k_z) \pm i / 2 \tau_0}, 
\eeq
where 
\beq
\label{eq:44}
\langle \br \alpha | n t k_y k_z \rangle = \frac{1}{\sqrt{L_z}} e^{i k_z z} z^t_{n \alpha}(k_z) \phi_{n k_y}(\br). 
\eeq

Evaluating $I(\bq, \Omega)$ in the general case is a formidable task, mostly due to the presence of the LL index sums. 
LLs with different $n$ will generally be mixed by impurity scattering, in which case analytical evaluation of Eq.~\eqref{eq:42} becomes impossible. 
However, we are primarily interested in transport along the direction of the applied magnetic field, i.e. the $z$-direction. 
If, in accordance with this, we set $\bq = q \hat z$, it is easy to see that only a single LL index sum remains in Eq.~\eqref{eq:42} after integration over the $x, y$ coordinates. 
In this case we obtain
\beqa
\label{eq:45}
I_{\alpha_1 \alpha_2, \alpha_3 \alpha_4} (q, \Omega)&=&\frac{\gamma^2}{2 \pi \ell_B^2 L_z} \sum_{n t t' k_z} 
\frac{z^{t}_{n \alpha_1}(k_z) \bar z^{t}_{n \alpha_3}(k_z)} {\Omega - \xi_{n t}(k_z + q/2) + i/ 2 \tau_0} \nonumber \\
&\times&\frac{z^{t'}_{n \alpha_4}(k_z) \bar z^{t'}_{n \alpha_2}(k_z)}{-\xi_{n t'}(k_z - q/2) - i/ 2 \tau_0}. 
\eeqa
The dependence of the eigenfunctions $z^t_{n \alpha}(k_z)$ on $q$ has been neglected in Eq.~\eqref{eq:45}, since the corresponding 
terms are subdominant in the limit $\epsilon_F \tau_0 \gg 1$ and $q v_F \tau_0 \ll 1$. 

At this point we need to explicitly separate out the part of the diffusion propagator that corresponds to hydrodynamic modes, i.e. modes with long relaxation times. 
On physical grounds, we expect only two such modes to be present in our system, corresponding to the diffusion of the vector $n_v = \sigma^0 \tau^0$ and the axial
$n_a = \sigma^z \tau^y$ charges. Note that the axial charge operator changes to $n_a =  \sigma^0 \tau^y$ after the canonical transformation Eq.~\eqref{eq:18.5}. 
The projection onto the vector-axial charge subspace is accomplished by the following transformation 
\beq
\label{eq:46}
{\cal D}^{-1}_{a b} = \frac{1}{4} (\sigma \tau)^a_{\alpha_2 \alpha_1} {\cal D}^{-1}_{\alpha_1 \alpha_2, \alpha_3 \alpha_4} (\sigma \tau)^b_{\alpha_3 \alpha_4}, 
\eeq
where $a, b$ refer to either the vector or the axial charge and summation over repeated indices is implied. 
The projected inverse diffusion propagator is a $2 \times 2$ matrix. 
Its diagonal components describe the independent transport of the vector and axial charge densities, while the off-diagonal components describe 
their coupling, induced by the applied magnetic field. 
Let us first evaluate the off-diagonal component (the two off-diagonal components are equal by reciprocity). 
We have
\beqa
\label{eq:47}
&&\cD^{-1}_{va}(\bq, \Omega) = - \frac{1}{4} (\sigma^0 \tau^0)_{\alpha_2 \alpha_1} I_{\alpha_1 \alpha_2, \alpha_3 \alpha_4}(\bq, \Omega) 
(\sigma^0 \tau^y)_{\alpha_3 \alpha_4} \nonumber \\
&=&-\frac{\gamma^2}{8 \pi \ell_B^2 L_z} \sum_{n t t' k_z} 
\frac{\langle z^{t'}_n(k_z) | z^t_n(k_z) \rangle}{\Omega - \xi_{n t}(k_z + q/2) + i/ 2 \tau_0} \nonumber \\
&\times&\frac{\langle z^t_n(k_z) | \tau^y | z^{t'}_n(k_z) \rangle}{-\xi_{n t'}(k_z - q/2) - i/ 2 \tau_0}. 
\eeqa
Since $\langle z^{t}_n(k_z) | z^{t'}_n(k_z)\rangle = \delta_{t t'}$, and 
\beq
\label{eq:48}
\langle z^t(k_z) | \tau^y | z^t(k_z) \rangle = - t \frac{\Delta_D \sin(k_z d)}{\Delta(k_z)}, 
\eeq
it is clear that LLs with $n \geq 1$ can not contribute to Eq.~\eqref{eq:47}, since their energies $\epsilon_{n t}(k_z)$ do not depend on the index $t$, 
which leads to an exact cancellation of contributions with $t = \pm$. 
The $n = 0$ LL, on the other hand, does contribute, since the corresponding eigenstate energies do depend on $t$, as seen in Eq.~\eqref{eq:23}. 
When $\epsilon_F > 0$, it is clear that only the $t = -$ lowest LL contributes to Eq.~\eqref{eq:47}, since only this LL crosses the Fermi energy. 
Using an identity
\beq
\label{eq:49}
A B = \frac{ B - A}{A^{-1} - B^{-1}}, 
\eeq
we have
\beqa
\label{eq:50}
&&\frac{1}{\Omega - \xi_{0 -}(k_z + q/2) + i/ 2 \tau_0} \frac{1}{-\xi_{0 -}(k_z - q/2) - i/ 2 \tau_0} \nonumber \\
&=&\frac{1}{\Omega - \xi_{0 -}(k_z+q/2) + \xi_{0 -}(k_z - q/2) + i/ \tau_0} \nonumber \\
&\times&\left[\frac{1}{-\xi_{0 -}(k_z - q/2) - i/ 2 \tau_0} - \frac{1}{\Omega - \xi_{0 -}(k_z + q/2) + i /2 \tau_0}\right] \nonumber \\
&\approx& \frac{2 \pi i \, \delta[\Delta(k_z) - \epsilon_F]}{\Omega - q \frac{d \Delta}{d k_z} + i/ \tau_0}. 
\eeqa
Substituting this into Eq.~\eqref{eq:47} and expanding to first order in $\Omega$ and $q$, we obtain
\beqa
\label{eq:51}
\cD^{-1}_{va}(\bq, \Omega)&=&\frac{\gamma^2 \tau_0}{8 \pi \ell_B^2} \int_{-\pi/d}^{\pi/d} d k_z \frac{1}{\Delta_S d} \frac{d \Delta}{d k_z} \delta[\Delta(k_z) - \epsilon_F] \nonumber \\
&\times&\left(1 + i \Omega \tau_0 - i q \frac{d \Delta}{d k_z} \tau_0 \right), 
\eeqa
where we have used 
\beq
\label{eq:52}
\frac{d \Delta}{d k_z} = - \frac{\Delta_S \Delta_D d \sin(k_z d)}{\Delta(k_z)} = - \Delta_S d \langle z^-(k_z) | \tau^y | z^-(k_z) \rangle. 
\eeq
Since $d \Delta/ d k_z$ is an odd function of $k_z$ with respect to the point $k_z = \pi/d$, it is clear that only the term, proportional to $q$, survives 
the integration over $k_z$ in Eq.~\eqref{eq:51}. 
Then we obtain
\beq
\label{eq:53}
\cD^{-1}_{va}(\bq, \Omega) = - i q \frac{\gamma^2 \tau_0^2}{4 \pi \ell_B^2 \Delta_S d} \left|\frac{d \Delta}{d k_z} \right|_{k_z = k_z^{\pm}}, 
\eeq
where $k^{\pm}_z = \pi/d \pm k_0$ are the two solutions of the equation $\Delta(k_z) = \epsilon_F$. 
Evaluating the lowest LL Fermi velocity $d \Delta/ d k_z$ explicitly, we obtain
\beq
\label{eq:54}
\left| \frac{d \Delta}{d k_z} \right|_{k_z = k_z^{\pm}} = \frac{d}{2}\sqrt{4 \Delta_S^2 - \epsilon_F^2} \approx \Delta_S d. 
\eeq
Then, using $\gamma^2 = 2/\pi g(\epsilon_F) \tau_0$, we finally obtain
\beq
\label{eq:54.5}
\cD^{-1}_{va}(\bq, \Omega) = \cD^{-1}_{av}(\bq, \Omega) = - i q \tau_0 \frac{e B}{2 \pi^2 g(\epsilon_F)} \equiv -i q \tau_0 \Gamma.
\eeq
The coefficient $\Gamma$ in Eq.~\eqref{eq:54.5} is a new transport coefficient, that describes the lowest LL-mediated coupling between the vector and the axial charge densities. 
This coupling may be regarded as being a consequence of chiral anomaly. 

The diagonal elements of the inverse diffusion propagator correspond to independent transport and relaxation of the vector 
and axial charge densities. 
These are nonzero in the absence of the magnetic field and we will thus evaluate them in the limit $B \rightarrow 0$, since we are 
interested in weak-field transport here. 
Accordingly, the contribution of the lowest LL is negligible in this case and we will ignore it. The limit of $B \rightarrow 0$ will 
be taken after summing the contributions of all the $n \geq 1$ LLs. 
One obtains
\beq
\label{eq:55}
\cD^{-1}_{vv}(\bq, \Omega) = 1 - \frac{\gamma^2 \tau_0}{4 \pi \ell_B^2} \sum_{n \geq 1} \int_{-\pi/d}^{\pi/d} d k_z \frac{\delta[\epsilon_n(k_z) - \epsilon_F]}
{1 - i \Omega \tau_0 + i q \tau_0 \frac{\Delta}{\epsilon_F}\frac{d \Delta}{d k_z}}.  
\eeq
The sum over the LL index $n$ may be done in the limit $B \rightarrow 0$ by converting the sum into an integral, just as was done when solving the SCBA equation above. 
Performing the integral and expanding to leading non vanishing order in $\Omega$ and $q$, one obtains
\beq
\label{eq:56}
\cD^{-1}_{vv}(\bq, \Omega) = - i \Omega \tau_0 + q^2 \tau_0^2 \left\langle \left(\frac{\Delta}{\epsilon_F} \frac{d \Delta}{d k_z}\right)^2 \right\rangle, 
\eeq
where the angular brackets denote average over the Fermi surface, defined as in Eq.~\eqref{eq:36}. 
Using
\beq
\label{eq:57}
\Delta \frac{d \Delta}{d k_z} = \frac{1}{2} \frac{d \Delta^2}{d k_z} = - \Delta_S \Delta_D \sin(k_z d), 
\eeq
the average is easily evaluated and we obtain
\beq
\label{eq:58}
\left\langle \left(\frac{\Delta}{\epsilon_F} \frac{d \Delta}{d k_z}\right)^2 \right\rangle \approx \frac{1}{3}(\Delta_S d)^2. 
\eeq
Defining the $z$-direction diffusion coefficient as $D = (\Delta_S d)^2 \tau_0 / 3$, we finally obtain
\beq
\label{eq:59}
\cD^{-1}_{vv}(\bq, \Omega) = - i \Omega \tau_0 + D q^2 \tau_0. 
\eeq
This has the expected form for the inverse diffusion propagator of a conserved quantity. Namely, the full diffusion propagator $\cD(\bq, \Omega)$ will exhibit a diffusion pole at 
$\Omega, q \rightarrow 0$ as a consequence of an exact conservation of the vector charge. 

Finally, we need to evaluate $\cD^{-1}_{aa}$. Here we expect that the diffusion pole will be absent due to a finite relaxation rate for the 
axial charge density, since it is not an exactly conserved quantity. 
We obtain
\beqa
\label{eq:60}
\cD^{-1}_{aa}(\bq, \Omega)&=&1- \frac{\gamma^2 \tau^0}{8 \pi \ell_B^2} \sum_{n \geq 1, t t'} \int_{-\pi/d}^{\pi/d} d k_z |\langle z^t_n(k_z) |\tau_y| z^{t'}_n(k_z) \rangle |^2 \nonumber \\
&\times&\frac{\delta[\epsilon_n(k_z) - \epsilon_F]}{1 - i \Omega \tau_0 + i q \tau_0 \frac{\Delta}{\epsilon_F}\frac{d \Delta}{d k_z}}, 
\eeqa
Evaluating the matrix element in Eq.~\eqref{eq:60}, one obtains
\beq
\label{eq:61}
\frac{1}{2} \sum_{t t'}  |\langle z^t_n(k_z) |\tau_y| z^{t'}_n(k_z) \rangle |^2 = 1 - \frac{\Delta^2(k_z) - \Delta_D^2 \sin^2(k_z d)}{\epsilon_n^2(k_z)}.
\eeq
Substituting this back into Eq.~\eqref{eq:60} and evaluating the sum over the LL index by converting it to an integral, as before, and expanding to leading 
non vanishing order in $\Omega$ and $q$, we get
\beqa
\label{eq:62}
&&\cD^{-1}_{aa}(\bq, \Omega) = \nonumber \\
&&1 -  \left \langle 1- \frac{\Delta^2(k_z) - \Delta_D^2 \sin^2(k_z d)}{\epsilon_F^2} \right \rangle (1 + i \Omega \tau_0) \nonumber \\
&+&q^2 \tau_0^2  \left \langle \left(1- \frac{\Delta^2(k_z) - \Delta_D^2 \sin^2(k_z d)}{\epsilon_F^2}\right) \left(\frac{\Delta}{\epsilon_F} \frac{d \Delta}{d k_z} \right)^2 \right \rangle, 
\nonumber \\
\eeqa
where the angular brackets again mean average over the Fermi surface.  
Evaluating the Fermi surface averages, assuming as before that $\epsilon_F/\Delta_S \ll 1$, we finally obtain
\beq
\label{eq:63}
\cD^{-1}_{aa}(\bq, \Omega) = - i \Omega \tau_0 + \frac{\tau_0}{\tau_a} + D q^2 \tau_0, 
\eeq
where 
\beq
\label{eq:64}
\frac{1}{\tau_a} = \frac{\epsilon_F^2}{20 \Delta_S^2 \tau_0}, 
\eeq
is the axial charge relaxation rate. Eq.~\eqref{eq:64} is one of the main results of this section. 
As expected, the axial charge is not exactly conserved, as the chiral symmetry is always explicitly violated in a real Dirac semimetal by 
nonlinearity of the band dispersion, which is necessarily present. 
However, since the band dispersion becomes more and more linear as the energy is reduced towards the Dirac point, the axial relaxation rate
tends to zero as the Fermi energy goes to zero faster than the momentum relaxation rate $1/\tau_0$, which of course also vanishes in the limit $\epsilon_F \rightarrow 0$ 
due to the vanishing density of states (This is true provided we neglect the influence of the magnetic field on the density of states.  
In principle, even in the limit $\epsilon_F \rightarrow 0$ there is a finite density of states, proportional to $B$. We ignore this in the weak-field limit). 
Thus, near the Dirac point $\tau_a \gg \tau_0$, which expresses the near-conservation of the 
axial charge due to the emergent low-energy chiral symmetry. As will be seen below, this is a necessary condition for a large negative magnetoresistance. 

Collecting all the matrix elements, we obtain the following result for the full inverse diffusion propagator, which describes coupled transport of the vector 
and axial charge densities
\beqa
\label{eq:65}
\cD^{-1}(q, \Omega) = \left(
\begin{array}{cc}
 -i \Omega \tau_0 + D q^2 \tau_0 & -i q \Gamma \tau_0 \\
 - i q \Gamma \tau_0 & -i \Omega \tau_0 + \tau_0/ \tau_a + D q^2 \tau_0
 \end{array}
 \right). \nonumber \\
\eeqa 
Viewing Eq.~\eqref{eq:65} as the inverse Green's function of the diffusion equation for the vector and axial charges and performing the inverse Fourier transform, 
we obtain the coupled diffusion equations
\beqa
\label{eq:66}
\frac{\partial n_v}{\partial t}&=&D \frac{\partial^2 n_v}{\partial z^2} + \Gamma \frac{\partial n_a}{\partial z}, \nonumber \\
\frac{\partial n_a}{\partial t}&=&D \frac{\partial^2 n_a}{\partial z^2} - \frac{n_a}{\tau_a} + \Gamma \frac{\partial n_v}{\partial z}. 
\eeqa
Since the vector charge is exactly conserved, the first equation must have the form of the continuity equation
\beq
\label{eq:67}
\frac{\partial n_v}{\partial t} = - \bnabla \cdot \bj_v,
\eeq
where $\bj_v$ is the vector current, i.e. current of the vector charge. 
This leads to the following explicit expression for the vector current
\beq
\label{eq:68}
j_v = - \frac{\sigma_0}{e} \frac{\partial \mu_v}{\partial z} - \frac{e^2 B}{2 \pi^2} \delta \mu_a.
\eeq
Here $\mu_v$ and $\mu_a$ are the vector and axial electrochemical potentials, $\sigma_0 = e^2 g(\epsilon_F) D$ is the zero-field Drude conductivity, and we have used 
$\delta n_{v, a} = g(\epsilon_F) \delta \mu_{v, a}$.
The first term in Eq.~\eqref{eq:68} is an ordinary current in response to a gradient of the electrochemical potential. 
The second term is a consequence of chiral anomaly and is an extra contribution to the current, proportional to the applied magnetic field and 
(nonequilibrium part of) the axial electrochemical potential. This is known as CME in the literature,~\cite{Kharzeev08} and this extra contribution is what leads to 
the anomalous negative longitudinal magnetoresistance. However, CME by itself is only one component of the experimentally observable effect, i.e. the negative magnetoresistance. 
The second crucial component, without which the effect is unobservable, is contained in the second of Eq.~\eqref{eq:66}. 
Namely, as will be seen shortly, the CME leads to observable magnetoresistance only if the axial charge relaxation rate $1/\tau_a$ is small, i.e. the axial charge is a nearly conserved quantity. 
As discussed above, this near-conservation of the axial charge is a characteristic feature of Dirac (and Weyl) semimetals, which becomes more and more precise as 
the Fermi energy is reduced towards the Dirac (or Weyl) point. 

To obtain the CME-related magnetoresistance explicitly, we now assume that there is a uniform steady state vector current density in the $z$-direction $j_v$, present in the system. 
The second of Eqs.~\eqref{eq:66} then gives
\beq
\label{eq:69}
\delta \mu_a = \Gamma \tau_a \frac{\partial \mu_v}{\partial z}. 
\eeq
Substituting this into the equation for the vector current Eq.~\eqref{eq:68}, we obtain the following expression for the total diagonal conductivity
\beq
\label{eq:70}
\sigma_{zz} = \sigma_0 + \frac{e^4 B^2 \tau_a}{4 \pi^4 g(\epsilon_F)}. 
\eeq
Thus CME is manifested as a positive longitudinal magnetoconductivity (or negative magnetoresistivity), proportional to $B^2$. 
Crucially, it is also proportional to $\tau_a$, and a large $\tau_a$ is thus necessary for this effect to be significant. 
The magnetoresistance is further enhanced when $\epsilon_F \rightarrow 0$ by vanishing of the density of states as $g(\epsilon_F) \sim \epsilon_F^2$. 
\section{Anomalous density response in a Weyl metal}
\label{sec:4}
In this section we will extend the theory of the anomaly-related negative magnetoresistance, presented above, to the case of Weyl metals, where 
the individual Weyl fermion components of the Dirac fermion are separated to distinct points in momentum space. 
A shorter account of this work has already been presented in Refs.~\onlinecite{Burkov14-3,Burkov15-1}. 
As the calculations are quite similar to the case of Dirac metals, described in the previous section, here we will only focus on the differences from 
the Dirac metal case and skip some of the details. 

In the context of our model Dirac semimetal, described by Eq.~\eqref{eq:18}, the separation of the Dirac fermion into Weyl fermions is most easily accomplished by adding a 
term $b \, \sigma^z$ to the Hamiltonian. 
Physically this term may arise from magnetized impurities, doped into the Dirac semimetal, or even from the Zeeman coupling to the applied magnetic field.    

The LL energy eigenvalues now have the form
\beq
\label{eq:71}
\epsilon_{na}(k_z) = s \sqrt{2 \omega_B^2 n + m_t^2(k_z)} \equiv s \epsilon_{n t}(k_z), \,\, n \geq 1, 
\eeq
while 
\beq
\label{eq:72}
\epsilon_{0 t}(k_z) = -m_t(k_z). 
\eeq
Here 
\beq
\label{eq:73}
m_t(k_z) = b + t \Delta(k_z). 
\eeq
Taking $b$ to be nonnegative, $m_-(k_z) $ vanishes at two points along the $z$-axis in momentum space, given by the two solutions 
of the equation
\beq
\label{eq:74}
\Delta(k_z) = b. 
\eeq
The solutions are $k_z^{\pm} = \pi/d \pm k_0$, where 
\beq
\label{eq:75}
k_0 = \frac{1}{d}\arccos\left(\frac{\Delta_S^2 + \Delta_D^2 - b^2}{2 \Delta_S \Delta_D}\right). 
\eeq
These correspond to the locations of the two Weyl nodes of opposite chirality on the $z$-axis in momentum space. 
The nodes exist as long as $b_{c1} < b < b_{c2}$ where $b_{c1} = |\Delta_S - \Delta_D|$ and $b_{c2} = \Delta_S + \Delta_D$. 
The eigenvectors are given by 
\beqa
\label{eq:76}
&&|v^{s t}_{n}(k_z) \rangle = \frac{1}{\sqrt{2}} \left(\sqrt{1 + s \frac{m_t(k_z)}{\epsilon_{n t}(k_z)}}, - i s \sqrt{1 - s \frac{m_t(k_z)}{\epsilon_{n t}(k_z)}} \right), \nonumber \\
&&|u^t(k_z) \rangle = \frac{1}{\sqrt{2}} \left(1, t \frac{\Delta_S + \Delta_D e^{- i k_z d}}{\Delta(k_z)} \right),
\eeqa
while the $n=0$ LL is polarized downwards, as before. 

The main difference from the Dirac metal case is that the Kramers degeneracy between the $t = \pm$ states is now broken by the spin splitting term $b \sigma^z$. 
When $b$ is sufficiently large (i.e. $b > \epsilon_F$), we may ignore the $t = +$ states entirely. 
Solving the SCBA equations as before, we obtain
\beq
\label{eq:76.5}
\frac{1}{\tau(k_z)} = \frac{1}{\tau_0} \left[1 + \frac{m_-(k_z) \langle m_- \rangle}{\epsilon_F^2} \right], 
\eeq
where $1/\tau_0 = \pi \gamma^2 g(\epsilon_F)$ and
\beq
\label{eq:77}
g(\epsilon_F) = \frac{1}{2 \pi \ell_B^2} \int_{-\pi/d}^{\pi/d} \frac{d k_z}{2 \pi} \sum_n \delta[\epsilon_{n -}(k_z) - \epsilon_F], 
\eeq
is the total density of states at Fermi energy. 
At the Weyl nodes $m_-(k_z)$ vanishes and changes sign. This implies that, for sufficiently small Fermi energy, such that the band dispersion 
in the $z$-direction may be assumed to be linear to a good approximation, the Fermi surface average $\langle m_-(k_z) \rangle$ will vanish. 
This property has a simple geometrical interpretation. The Weyl nodes may be thought of as monopole sources of Berry curvature, whose 
$z$-component is proportional to $m_-(k_z)$. This clearly averages to zero when integrated over the volume, enclosed by a sufficiently small Fermi 
surface sheet, containing the node.~\cite{Burkov14-2}
Assuming this to be the case, we obtain
\beq
\label{eq:78}
\frac{1}{\tau(k_z)} \approx \frac{1}{\tau_0}. 
\eeq
Note that since the densities of states in Eqs.~\eqref{eq:35} and \eqref{eq:77} are essentially identical in the limit of small Fermi energy (the two-fold Kramers degeneracy in the Dirac metal case is replaced by two identical Fermi surfaces, enclosing the Weyl nodes, in the Weyl metal case), the impurity scattering rate in the Weyl metal is twice as large. 
This is easy to understand physically and follows simply from the near orthogonality of the $|u^{\pm}(k_z)\rangle$ eigenstates at small momentum difference, i.e. 
$\langle u^t(k_z) | u^{t'}(k_z') \rangle \approx \delta_{t t'}$ when $|k_z - k_z'| d \ll 1$, which means that scattering between the two components of the Kramers doublet is 
suppressed in the Dirac metal. This suppression disappears in the Weyl metal case, since in this case the two Fermi surfaces arise from states in the same $t = -$ band. 

The evaluation of the diffusion propagator goes along exactly the same lines as before. The only difference is that only the $t= -$ states contribute, i.e. we have
\beqa
\label{eq:79}
I_{\alpha_1 \alpha_2, \alpha_3 \alpha_4} (q, \Omega)&=&\frac{\gamma^2}{2 \pi \ell_B^2 L_z} \sum_{n k_z} 
\frac{z^{-}_{n \alpha_1}(k_z) \bar z^{-}_{n \alpha_3}(k_z)} {\Omega - \xi_{n -}(k_z + q/2) + i/ 2 \tau_0} \nonumber \\
&\times&\frac{z^{-}_{n \alpha_4}(k_z) \bar z^{-}_{n \alpha_2}(k_z)}{-\xi_{n -}(k_z - q/2) - i/ 2 \tau_0}. 
\eeqa
For the same reason, the projection of the full diffusion propagator onto the vector and axial charge subspace differs by a factor of $1/2$ from the Dirac semimetal case
\beq
\label{eq:80}
{\cal D}^{-1}_{a b} = \frac{1}{2} (\sigma \tau)^a_{\alpha_2 \alpha_1} {\cal D}^{-1}_{\alpha_1 \alpha_2, \alpha_3 \alpha_4} (\sigma \tau)^b_{\alpha_3 \alpha_4}, 
\eeq
The vector to axial charge coupling term arises, as before, entirely from the contribution of the $n=0$ LL. We obtain
\beq
\label{eq:81}
\cD^{-1}_{va}(\bq, \Omega) = - i q \tau_0 \frac{1}{2 \pi \ell_B^2 g(\epsilon_F) \Delta_S d} \left|\frac{d \Delta}{d k_z} \right|_{k_z = k_z^{\pm}}, 
\eeq
where $k_z^{\pm}$ are now solutions of the equation
\beq
\label{eq:82}
\Delta(k_z) = b + \epsilon_F, 
\eeq
which determines the points at which the Fermi energy intersects the $n = 0$, $t = -$ LL. 
One obtains
\beq
\label{eq:83}
\left|\frac{d \Delta}{d k_z} \right|_{k_z = k_z^{\pm}} = \frac{d}{2(b + \epsilon_F)} \sqrt{[(b + \epsilon_F)^2 - b_{c1}^2][b_{c2}^2 - (b + \epsilon_F)^2]}. 
\eeq
Assuming $b_{c1} \ll b + \epsilon_F \ll b_{c2}$, which implies that the Weyl node splitting and the Fermi energy are such that the band dispersion at the Fermi 
level may be taken to be linear, one obtains
\beq
\label{eq:84}
\left|\frac{d \Delta}{d k_z} \right|_{k_z = k_z^{\pm}} \approx \frac{\Delta_S + \Delta_D}{2} d \approx \Delta_S d.
\eeq
This gives 
\beq
\label{eq:85}
\cD^{-1}_{va}(\bq, \Omega) = - i q \tau_0 \frac{e B}{2 \pi^2 g(\epsilon_F)} \equiv - i q \tau_0 \Gamma, 
\eeq
i.e. an identical result to what we obtained before in the case of the Dirac metal. 

The diagonal elements of the inverse diffusion propagator are also evaluated in exactly the same way as in the case of the Dirac metal. 
The form of the expression for the vector charge part of the propagator is, as before, constrained by the vector charge conservation
\beq
\label{eq:86}
\cD^{-1}_{vv}(\bq, \Omega) = -i \Omega \tau_0 + D q^2 \tau_0, 
\eeq
where the diffusion coefficient is given by
\beq
\label{eq:87}
D = \tilde v_F^2 \tau_0 \left \langle \frac{m_-^2(k_z)}{\epsilon_F^2} \right \rangle, 
\eeq
which is identical to the corresponding result in the Dirac metal case but with $\Delta(k_z)$ replaced by $m_-(k_z)$. 
The Fermi velocity in Eq.~\eqref{eq:87} is $\tilde v_F(k_z) = |d \Delta/d k_z|$, evaluated at the Weyl node locations, which is given by
\beq
\label{eq:88}
\tilde v_F = \frac{d}{2 b} \sqrt{(b^2 - b_{c1}^2)(b_{c2}^2 - b^2)}.
\eeq
The average of $m_-^2(k_z)$ over the Fermi surface may be easily evaluated in the limit of small Fermi energy, which allows one to expand $m_-(k_z)$ to leading 
order in deviation of $k_z$ from the Weyl node locations. 
In this case one obtains
\beq
\label{eq:89}
D \approx \frac{1}{3} \tilde v_F^2 \tau_0, 
\eeq
i.e. again identical to the corresponding Dirac metal result, obtained in Section~\ref{sec:3}. 

Finally, for the axial charge block of the inverse diffusion propagator we obtain the following expression
\beq
\label{eq:90}
\cD^{-1}_{aa}(\bq, \Omega) = -i \Omega \tau_0 + \frac{\tau_0}{\tau_a} + D q^2 \tau_0,  
\eeq
where the axial charge relaxation rate is now given by
\beq
\label{eq:91}
\frac{\tau_0}{\tau_a} = \frac{1 - (\tilde v_F/\Delta_S d)^2}{(\tilde v_F/ \Delta_S d)^2}. 
\eeq
This expression for the axial charge relaxation rate represents the most significant difference of the Weyl metal case from the Dirac metal case. 
This equation shows, in particular, that in a Weyl metal the (dimensionless) axial charge relaxation rate is essentially always finite, even in the
limit $\epsilon_F \rightarrow 0$. It may still be expected to be small, which is easily seen explicitly in the limit when $b_{c1} \ll b \ll b_{c2}$, and $\epsilon_F \ll b$. 
In this case we obtain 
\beq
\label{eq:92}
\frac{1}{\tau_a} \approx \frac{b^2}{4 \Delta_S^2 \tau_0}, 
\eeq
i.e. the axial charge relaxation increases quadratically with the spin-splitting parameter $b$ as the Weyl nodes get split and separated out of the parent Dirac metal state.
The diffusion equations themselves and the magnetoresistance formula are identical to the Dirac semimetal case, with the axial charge relaxation rate given by Eq.~\eqref{eq:92}, so 
we will not repeat them explicitly here. 

Before we conclude this section, we would like to point out an important caveat, which applies to the results of this section. 
Namely, all of the results above are only applicable if the condition 
\beq
\label{eq:93}
b_{c1} \ll b \ll b_{c2}, 
\eeq
is satisfied. 
An implicit assumption here is that Weyl semimetal is obtained from a parent state, which is nearly a Dirac semimetal (hence the small $b_{c1} = |\Delta_S - \Delta_D|$), 
and we are far from the transition out of the Weyl semimetal state in the large $b$ limit, i.e. when $b = b_{c2}$. 
Only if the condition Eq.~\eqref{eq:93} is satisfied may we expect to get negative magnetoresistance, quadratic in the magnetic field in the general case. 
Otherwise, in magnetic Weyl semimetals, linear magnetoresistance, which is allowed by symmetry, will dominate the quadratic one at small fields. 
Under the condition Eq.~\eqref{eq:93}, the linear terms in magnetoresistance are $O(b/b_{c2})$ and may thus be ignored. 
Linear magnetoresistance is strictly absent by symmetry, of course, in the case of noncentrosymmetric Weyl semimetals. 
In this case the above results apply without restriction. 
\section{Discussion and conclusions}
\label{sec:5}
In this paper we have developed a theory of anomaly-related weak-field quadratic negative longitudinal magnetoresistance in Dirac and Weyl metals. 
An important issue is how to differentiate this novel magnetoresistance from other possible contributions, which are more mundane in origin. 
In fact, longitudinal magnetoresistance, which is what we are interested in here, in never entirely mundane. 
The reason is that, from the simplest Drude theory viewpoint, the only possible source of magnetoresistance is the Lorentz force, which is of course absent for electrons, propagating 
along the direction of the field. 
Drude theory thus predicts that longitudinal magnetoresistance is always absent, which is not the case: there are many examples of materials, exhibiting it, even at low fields. 
Several possible sources of longitudinal magnetoresistance have been identified over the years,~\cite{Sondheimer62,Stroud76,Miller96} but perhaps the most universal source, related purely to the intrinsic properties of the 
electronic structure, was described recently by Pal and Maslov.~\cite{Maslov10} They have shown that longitudinal weak-field magnetoresistance arises necessarily when the shape of the Fermi surface exhibits certain types of angular anisotropy with respect to the direction of the magnetic field. 
This mechanism gives positive magnetoresistance, increasing quadratically with the magnetic field at low fields. 
For the anomaly-related negative magnetoresistance to be observable, it needs to be larger than this Fermi surface anisotropy-driven magnetoresistance. 
From this viewpoint, the Dirac metal case seems to be the best: one may expect both a weak anisotropy and a large axial relaxation time in this case. 
Weyl metal case with either a very large separation between the nodes, i.e. separation approaching the size of the first BZ, or a very small separation (unless it arises from a 
parent zero-gap Dirac semimetal), are both problematic, since the axial charge relaxation time may be expected to be small in these cases. 

It is also important to remember that the theory, presented in this paper, applies only in the semiclassical limit, i.e. $\omega_B/ \epsilon_F \ll 1$. 
In the opposite, ultraquantum limit, one may expect a linear negative magnetoresistance,~\cite{Nielsen83,Aji12} which may be obtained from Eq.~\eqref{eq:70}
by substituting the lowest LL density of states $g(\epsilon_F) \sim B$. 
It is also possible to have quadratic magnetoresistance of both signs in this limit, arising from the combined action of chiral anomaly and field-induced modification of the density of states.~\cite{Goswami15}
One hopes that the current experiments~\cite{Kharzeev14,Hasan15-2,Ong15-1,Ong15-2,Fang15-3,Li15,Zhang15} are in the semiclassical limit, as the quadratic magnetoresistance is observed at low fields, but the magnitude of the ratio $\omega_B/\epsilon_F$ is at the moment uncertain in these experiments. 

Finally, in the theory, developed in the paper, a particular model of the impurity scattering was assumed: weak point-like Gaussian-distributed scatterers.
This assumption was made primarily for computational convenience: the ladder sum for the diffusion propagator, Eq.~\eqref{eq:41}, may only be calculated straightforwardly in this case, which is a limitation of the present approach. 
In many cases, a more physically-realistic model should involve Coulomb impurities.
In this case one might in fact expect that the axial relaxation rate should be even smaller, at least in the Weyl semimetal case, as the finite-momentum scattering, 
necessary to scatter electrons between the nodes, will be suppressed. However, a detailed calculation of this would certainly be helpful.
What happens to the magnetoresistance in the strong disorder limit~\cite{Brouwer14-1,Gurarie15-1,Gurarie15-2,Brouwer15-1} is also an important and experimentally-relevant issue, 
worthy of a thorough study. 

In conclusion, we have presented a theory of chiral anomaly-driven negative quadratic longitudinal magnetoresistance in Dirac and Weyl metals in the weak magnetic 
field regime. 
We have demonstrated that this effect has two crucial ingredients. One is the coupling between the vector and the axial charge density, proportional to the magnetic field, 
which arises from the chiral lowest Landau level, or the nontrivial Berry curvature of the band eigenstates. This coupling will in principle exist in any system with a nonzero 
Berry curvature 
and is in this sense not specific to Dirac or Weyl metals, although the transport coefficient, describing such a coupling, has a universal value of $e B/ 2 \pi^2$ in Dirac or Weyl metals only. 
The second ingredient is the near-conservation of the axial charge density, manifesting in large axial charge relaxation time $\tau_a/\tau_0 \gg 1$. 
This property is specific to Dirac and Weyl metals only and is necessary for the negative quadratic magnetoresistance to be observable.

\begin{acknowledgments}
Financial support was provided by Natural Sciences and Engineering Research Council (NSERC) of Canada. 
\end{acknowledgments}
\bibliography{references}

\end{document}